\documentclass[twocolumn]{revtex4}

\usepackage{dcolumn}
\usepackage{bm}

\begin{document}

\title {A progenitor model of Cosmic Ray knee} 
\author{Biplab Bijay} 
\email{biplabbijay@rediffmail.com}
\author{Arunava Bhadra} 
\email{aru\_bhadra@yahoo.com}
\affiliation{High Energy \& Cosmic Ray Research Centre, University of North Bengal, Siliguri, WB 734013 INDIA}

\begin{abstract} Primary energy spectrum of cosmic rays exhibits a knee at about $3$ PeV where a change in the spectral index occurs. Despite many efforts the origin of such a feature of the spectrum is not satisfactorily solved yet. Here it is proposed that the steepening of the spectrum beyond the knee may be a consequence of mass distribution of progenitor of cosmic ray source. The proposed speculative model can account all the major observed features about cosmic rays without invoking any fine tuning to match flux or spectra at any energy point. The prediction of the proposed model regarding primary composition scenario beyond the knee is quite different from most of the prevailing models of the knee and thereby can be discriminated from precise experimental measurement of the primary composition.
\end{abstract}

\keywords{cosmic rays:  acceleration of particles: black hole physics}

\maketitle

%
%
\section{Introduction}
Ever since their discovery more than a hundred years back now, the origin of cosmic rays has been one of the central question of physics. But despite many efforts so far there is no consistent and complete model of the origin of cosmic rays. \\
The energy spectrum of cosmic rays provides important clues about their origin. The most intriguing feature of the energy spectrum is that though it extends a wide range of energies, from sub GeV to at least $3 \times 10^{20}$ eV (the highest energy observed so far), it can be well represented by a steeply falling power law for energies above the solar modulated one. However, the spectrum has a knee around $3$ PeV where it steepens sharply as discovered more than half a century ago by Kulikov and Khristiansen of the Moscow State University \cite{kul59}. The spectrum also has an ankle at an energy about $3$ EeV where it flattens again to its pre-knee slope. It is relatively easier to interpret the flattening of the spectrum above the ankle as the eventual superseding of a harder cosmic ray component which is sub-dominant at lower energies. In contrast the feature of knee is more difficult to explain. The existence of the knee in the spectrum is definitely an important imprint of the true model of origin of cosmic rays and hence a proper explanation of the knee is expected to throw light on the problem of cosmic ray origin. \\
Several mechanisms have been proposed so far to explain the knee. Shortly after the discovery of the knee, this spectral feature was interpreted as an effect of the reduced efficiency of galactic magnetic field to confine the cosmic ray particles with energies above the knee within galaxy \cite{gin64, wdo84, ptu93, can02, gia14}. Since the magnetic rigidity of a particle is proportional to its atomic number (Z), cosmic ray protons should start escaping first and hence the observed knee is the proton knee as per this model. \\
The knee has also been explained based on acceleration mechanism \cite{fic86, jok86, bie93, ber99, sta93, kob02}. For reasons of the power required to maintain the observed cosmic ray energy density, it is widely accepted that cosmic rays up to the ankle energy are of galactic origin whereas those having energies above this energy are extragalactic though there are also suggestions for lower transitional energies \cite{bla14, ama14,alo12}. Among the galactic sources supernova remnants (SNRs) satisfy the energy budget of cosmic rays. The power law behavior of the energy spectrum on the other hand suggests that cosmic rays are most probably energized by diffusive shock acceleration. The maximum energy that a charged particle can gain by diffusive shock acceleration is proportional to Z. The knee has been assigned in this model as the maximum energy that protons can have under diffusive shock acceleration in SNRs. \\
A critical analysis of the world data on energy spectrum suggests that the knee is very sharp, the spectral slop changes rather abruptly at the knee position \cite{erl97}. In contrast, the above mentioned rigidity dependent explanations of the knee predict a smooth change in the spectral slope at the knee because of sum of the contributions of different atomic nuclei having cut-offs at different energies (depending on Z values). To accommodate the sharp knee feature, a few proposals were advanced. In the single source model the dominant contribution of the cosmic ray flux at the knee is by a nearby source \cite{erl97, bha05, erl11, ter14} which is superimposed on a galactic modulated component in which spectral slop is changing smoothly with energy. In another model the sharp knee is explained in terms of cosmic ray acceleration by a variety of supernovae \cite{sve04, sve03}. The later proposal relies on the fact that the explosion energy of all the supernovae are not the same. The sharp knee also could be due to interaction of cosmic ray particles from a pulsar with radiation from the parent supernova remnant \cite{hu09}. \\         
The mass composition of cosmic rays will be heavier beyond the knee if the knee is a proton knee. Several Extensive Air Shower (EAS) measurements (till now the study of cosmic rays above 1 PeV is of indirect nature via the EAS observations) have been made to determine the mass composition of cosmic rays in the concerned energy region but the measurements have not yielded mutually consistent results yet due to the weak mass resolution of the measured shower observables \cite{hau11}. Most of the findings \cite{nav98, gla99, aar13, fom96} based on electron content relative to muon content (or {\itshape vice versa}) in EAS suggest that composition becomes heavier with energy beyond the knee though Haverah Park and few other observations (particularly underground muon telescopes) \cite{bla98, bla95, dan95, cha98, agl90, ahl92, kas97, lon95, bak99} found opposite trend of mass composition. Mass composition estimated from measurement of the depth of shower maximum through observation of Cerenkov \cite{boo97, swo01, fow01, che04, kar95, arq00, dic99, efi91} or fluorescence radiation \cite{abr10, abb08, abb04, tsu11, jui12} on the other hand suggest lighter mass composition beyond the knee differing from those obtained with muon to electron content ratio \cite{hau11,hor13, bha05a}. The mass composition picture of primary cosmic rays is thus still inconclusive in the PeV and higher energy region. \\
Considering the possibility that mass composition may become lighter beyond the knee an alternative explanation of knee was suggested based on nuclear photo-disintegration at the sources \cite{hil79, kar93, can02a}. In this scenario heavier components of cosmic rays, particularly Fe nuclei undergo nuclear photo-disintegration in interaction with the radiation field of the source so that flux of heavier nuclei decreases with energy beyond the knee whereas protons loose energy by photo-meson production. \\
A major problem with the standard scenario of diffusive shock acceleration of cosmic rays at SNRs is that a cosmic ray particle hardly attain the knee energy under this SNR shock acceleration scenario. Such a problem can be overcome in the Cannonball model \cite{dar99, pla02, dar05, der05} in which masses of baryonic plasma or the so called cannonballs, ejected ultra-relativistically in bipolar supernova explosions, are considered as universal sources of hadronic galactic cosmic rays. In this model the knee corresponds to maximum energy gained by nuclei by elastic magnetic scattering of ambient ISM particles in the Cannonball while re-acceleration of cosmic rays by Cannonballs of other supernova explosions causes the extra steepness above the knee.\\
There is also proposal of explaining the knee based on a change in the characteristics of high energy interactions \cite{nik00}. In this model the knee is not a feature of the primary cosmic ray energy spectrum itself, but is caused by change in high-energy interaction characteristics, either producing a new type of a heavy particle unseen by air shower experiments, or an abrupt increase in the multiplicity of produced particles. However, this proposal is ruled out at present as the assumed interaction features have not been observed in the LHC experiment. \\   
None of the prevailing models of knee are free from problems. If the knee corresponds to break in proton spectrum, either because of it is the maximum energy to which proton can accelerate in a galactic cosmic ray source or due to begin of proton leakage from the galaxy at this energy with or without modifications for the sharp knee, then there should be a Fe knee around $10^{17}$ eV. Hence a special variety of supernovae or some other type of galactic or extra-galactic source has to be invoked as generator of cosmic rays between $\sim 10^{17}$ eV and the ankle or galactic extra-galactic transition should occur at around $10^{17}$ eV. The problem with the later proposal is that it requires fine-tuning to match both the flux and the energy at the point of taking over. The cannonball model also suffers the same fine tuning problem at the knee energy. There are other problems such as lower than expected observed gamma ray fluxes from SNRs. The dilemma of the knee thus still continues.\\
The viable sources of cosmic rays include SNRs, pulsars, Gamma Ray Bursts (GRBs), Active Galactic Nuclei (AGN) etc. Whatever may be the sources, there is little doubt that they are products of stellar evolution process. And an interesting fact is that the zero age mass spectrum (ZAMS) of stars also exhibits power law behavior \cite{sal55, kro02,mas95}. This immediately suggests that the cosmic ray energy spectrum might have some connection with the mass distribution of progenitor of their sources. In the present work we explore the idea and propose a model for the cosmic ray origin in which the knee of the primary cosmic ray energy spectrum at $\sim 3$ PeV is a consequence of mass distribution of progenitor of cosmic ray sources. The proposed model is free from any fine tuning problem and it also overcomes the issue of maximum attainable energy.  \\
The organization of the article is as follows. The model proposed in this work is presented in the next section. The outcome of the present model is discussed in section 3. The results of the model are compared with observations in section 4. Finally the results are concluded in section 5. \\

\section{The proposed model}
Here we propose a model of origin of cosmic rays in which there is a single class of major cosmic ray sources in the galaxy.\\
The basic conjectures of the present model are the followings: \\
1) Cosmic rays at least up to the ankle energy are produced either in gravitational explosions (core collapse) of  massive stars those lead to formation of black holes rather than neutron stars/white dwarf or in accretion onto black holes. No other type of galactic or extra-galactic source dominates at least up to the ankle energy. Here we have not identified the source. The probable candidate sources of cosmic rays include hypernova, Active Galactic Nuclei (AGNs) and Gamma Ray Bursts (GRBs). \\
2) Particles are accelerated by expanding shock waves up to a maximum energy $E_{max}$. The maximum attainable energy $E_{max}$ is, however, not the same for all the sources (of same kind) but depending on energy released in explosion/accretion, it has a range. The minimum $E_{max}$ that is possible for the cosmic ray sources is equal to the knee energy. We shall argue in the following section that the correspondence of minimum $E_{max}$  with the knee energy is quite plausible and suggestive. \\    
The observed cosmic ray luminosity demands that the cosmic ray sources must be energetically very powerful and are most likely to be powered by gravitational energy. The gravitational collapse that ultimately leads to formation of black hole or accretion onto black hole is expected to release maximum gravitational energy. This is the reason for considering the first conjecture. The maximum energy that a cosmic ray particle can attain in shock acceleration usually depends on the explosion energy. Since black hole has no limiting mass, energy released in black hole formation should varry with progenitor mass and hence the maximum attainable energy of cosmic ray particles are expected to be varry rather than having a fixed value. Essentially this is the logic behind the second conjecture. 

\subsection{The Progenitor connection}
Perhaps occurrence of relativistic shock and non-relativistic shock depends whether a black hole (BH) or a neutron star (NS) is formed in the stellar evolution processes. Through stellar core collapse progenitor stars with $M< 20 M_{\odot}$ are supposed to give rise to a neutron star or white dwarf whereas stars more massive than $20$ to $25 M_{\odot}$ form a black hole \cite{fry99, fry00, fry03} though such end point fate also depends on metallicity \cite{heg03}. The formation of white dwarf or neutron star is usually associated with supernova explosion. The masses of white dwarfs and neutron stars have to be  within the Chandrasekhar limit and Openheimer-Volkof limit respectively. Consequently the energy released in all ordinary supernova explosions are nearly the same. Since black hole has no such upper mass limit the energy released in core collapse of massive stars leading to black holes should depend on the mass of the progenitor star.   

The gravitational collapse of massive stars to black holes involves some complex, still poorly understood aspects of stellar physics. In the collapsar mechanism \cite{woo93}, a black hole is formed when the collapse of a massive star fails to produce a strong supernova explosion, leading to ultimate collapse into a black hole. If the stellar material falling back and accreting onto the black hole has sufficient angular momentum, it can hang up, forming a disk. This disk, by neutrino annihilation or magnetic fields, is thought to produce the jets which finally results in Active Galactic Nuclei (AGN) or hypernova. \\
In the gravitational collapse of a spherical mass distribution of rest mass M leading to formation of black hole, the maximum energy of extraction out of the collapse will be \cite{ruf03, chr71}.      
\begin{equation}
E_{max}^{collapse} =M c^{2}/2
\end{equation}

During the final stages of stellar evolution, a massive star losses a significant amount of mass. But if a black hole is formed stellar material likely to fall back and accreting onto the black hole \cite{woo93}. The mass of the final produced black hole is thus expected to increase linearly with the mass of progenitor.

Instead of collapse and resulting explosion, large amount of energy also can be released through accretion process. The Eddington luminosity limit, the maximum steady-state luminosity that can be produced is given by $L_{ed}=4\pi G M m_{p} c/\sigma_{\tau}$ where M is the mass of the black hole, $m_{p}$ is the proton mass and $\sigma_{\tau}$ is the Thomson cross section. The luminosity is thus also proportional to the mass of the black hole.

During the final stages of stellar evolution, a massive star losses a significant amount of mass. But if a black hole is formed stellar material likely to fall back and accreting onto the black hole \cite{woo93}. The mass of the final produced black hole is thus expected to increase linearly with the mass of progenitor, and hence the distribution of released energy is expected to follow the mass distribution of progenitors.

\section{4. Outcomes of the proposed model}

We shall now find out the outcomes of the proposed model regarding the main cosmic ray observables such as luminosity, maximum attainable energy, energy spectrum and nuclear composition. 

\subsection{The cosmic Ray Luminosity:}

The average energy released in BH formation should be around $5 \times 10^{53}$ ergs as per the equation (1) which is more than two order higher than that released in supernova explosion. Stars more massive than $20$ to $25 M_{\odot}$ usually form a black hole. The rate of stars having $M>20 M_{\odot}$ is $2 \times 10^{-3}\; yr^{-1}$. However, not all massive stars will end up as black holes. If we denote the probability of BH formation for a star massive than $20 \; M_{\odot}$  is $\rho_{BH}$, the total energy released in BH production during the cosmic ray confinement period of about $10^{6}$ years in the galaxy is about $\rho_{BH} 10^{57}$ ergs which yields a luminosity $3\rho_{BH} \zeta \times 10^{43}$ ergs/s, where $\zeta$ is the efficiency of conversion of explosion energy into cosmic ray energy. Typically $\zeta$ ranges from 0.01 to 0.1 whereas $\rho_{BH}$ may be taken as 0.5 \cite{cla15}.

\subsection{The Maximum attainable Energy:}

The maximum energy that a particle with charge Ze can be attained in a bulk magnetized flow on a scale $R_{s}$, with velocity $c \beta_{s}$ and magnetic field $B$ is \cite{hil84} 
\begin{equation}
E_{max}= ZeB\Gamma_{s} \beta_{s} R_{s}
\end{equation}
where $\Gamma_{s}$ is the Lorentz factor of the relativistic shock wave. This value of $E_{max}$ is a factor $\Gamma_{s}$ larger than that obtained from the Hillas condition. In a BH formation scenario, a fraction of all kinetic energy carries a debris ejected with the largest Lorentz factor thereby generating gamma ray emission in the form of burst, but the bulk of ejecta is less relativistic or even sub-relativistic. Note that if $\sim 10 M_{\odot}$ is given $\sim 10^{54}$ ergs then typical velocity of the mass would be $10^{10}$ cm i.e. $c/3$. GRBs are likely to occur in BH formation collapse and a hint on typical values of $\Gamma_{s}$ may be found from GRBs. The GRB observations suggest minimum $\Gamma_{s}$ of the burst is few tens \cite{rac11, lit01, zou11}. Therefore, the minimum $E_{max}$ for BH producing explosion should be of few PeV. \\

Let us consider a more rigorous description. In the standard scenario the acceleration of cosmic rays occurs at (non-relativistic) shocks of isolated supernova remnants (SNRs). The maximum energy that can be attained by a cosmic ray particle in an ordinary SNR when the remnant is passing through a medium of density $N_{H} \; cm^{-3}$ is  \cite{fic86, bie93, ber99}
\begin{eqnarray}
E_{max}\simeq 4 \times 10^{5} Z \left( \frac{E_{SN}}{ 10^{51} \; erg} \right)^{1/2} \left( \frac{M_{ej}}{ 10 M_{\odot}} \right)^{-1/6} \\ \nonumber
\left( \frac{N_{H}}{ 3 \times 10^{-3} \; cm^{-3}} \right)^{-1/3}\left( \frac{B_{o}}{ 3 \mu G} \right) \; GeV 
\end{eqnarray}
which is falling short of the knee by about one order of magnitude. Energy released in BH formation explosions is at least two order higher than that in SN explosion. Moreover, as stated before, for relativistic shock acceleration $E_{max}$ will be a factor $\Gamma_{s}$ higher. Hence the minimum $E_{max}$ for BH producing explosion should be of few PeV.  \\
An important question for such a BH formation explosion origin of cosmic rays is whether or not $E_{max}$ could reach the ankle energy. Unlike almost constant energy released in SN explosions, energy output in such a scenario varies and it may go at least up to 2 order high from its minimum value. And such high energy events are expected to occur in more rarefied medium. Hence it is very likely that the maximum $E_{max}$ will exceed the ankle energy.\\
Interestingly for AGN minimum $E_{max}$ is about 3 PeV \cite{ste91} which is the knee energy and the maximum $E_{max}$ can be many order higher than that owing to the wide range of luminosities of AGNs.\\

\subsection{Energy Spectrum:}
In the proposed model cosmic rays are accelerated in diffusive relativistic shock acceleration. The energy spectrum of accelerated particles in each source is, therefore, given by a power law

\begin{equation}
\frac{dn}{dE} = A E^{-\gamma} 
\end{equation}

with $\gamma$ is around $2.2$, and $A$ is the normalization constant 
\begin{equation}
A \equiv \frac{\epsilon}{(\gamma-2)(E_{min}^{-\gamma+2}-E_{max}^{-\gamma+2})}
\end{equation}
where $E_{min}$ and $E_{max}$ are respectively the minimum and maximum attainable energies of cosmic ray particles in the source.  \\
All the sources do not have the same $E_{max}$. Above the minimum possible $E_{max}$, which we denote as $E_{max}^{min}$, the spectrum will be modified due to the distribution of $E_{max}$. To get the spectrum beyond the $E_{max}^{min}$ we need to obtain maximum energy distributions of the cosmic ray sources from the mass distribution of their progenitors. The calculation involves a sequence of steps. Using the expression for explosion energy as function of progenitor mass as obtained in the previous section, we convolve the resulting explosion energy-progenitor mass relation with the initial mass function of the progenitors to obtain explosion energy distribution. Subsequently using the relation of maximum energy that a cosmic ray particle may attain in relativistic shock acceleration process with explosion energy, we derive the maximum energy distribution for main cosmic ray sources. Using such distribution we obtain the energy spectrum of cosmic rays beyond the $E_{max}^{min}$. \\
The stellar initial mass function, or distribution of masses with which stars are formed can be represented by a declining power law 
\begin{equation}
\frac{dn}{dM} \propto  M^{-\alpha} 
\end{equation}
with the universal (Salpeter) value of the exponent $\alpha = -2.35$ over the whole mass range above $3 M_{\odot}$ \cite{sal55, kro02,mas95}. Since explosion energy ($\epsilon$) scales linearly with $M$, the expected explosion energy distribution of massive progenitor stars, is also represented by $\frac{dn}{d\epsilon} \propto  \epsilon^{-\alpha}$. \\
The Lorentz factor of relativistic shock is nearly equal to the initial Lorentz factor of the jet i.e. $\Gamma_{s} \sim \gamma_{o}$. The relativistic shock waves must carry a significant fraction of the explosion energy which subsequently convert to energies of cosmic rays. Hence $\Gamma_{s}$ should be proportional to explosion energy. On the other hand $E_{max}$ is also proportional to $\Gamma_{s}$. So for the proposed model, $E_{max} \propto \epsilon $. Thus we have 
\begin{equation}
\frac{dn}{dE_{max}} \propto  E_{max}^{-\alpha} 
\end{equation}. 
Therefore, the number of sources having $E_{max} \ge E$ is $j(E_{max} \ge E) \propto E_{max}^{-\alpha +1}$. As the minimum $E_{max}$ of a source is equal to $E_{max}^{min}$ all such sources will contribute to cosmic ray flux when cosmic ray energy is below or equal $E_{max}^{min}$. However, for energies above the $E_{max}^{min}$ ($E>E_{max}^{min}$) only sources having $E_{max} \ge E$ will contribute. The resultant cosmic ray spectrum above the $E_{max}^{min}$ will be 
\begin{eqnarray}
\frac{dn}{dE} &=&\int_{E} \frac{dn}{dE_{max}} A E^{-\gamma} dE_{max}  \\ \nonumber
&&              \propto E^{-\gamma-\alpha+2} 
\end{eqnarray}
Therefore beyond the $E_{max}^{min}$ the spectrum should be steepen by $0.35$ in spectral index as observed. Note that the difference in power of energy by one between the above equation and the Eq.(3) of Kachelriess and Semikoz \cite{kac05}, where power law distribution of maximum attainable energy of sources was assumed, is due to the fact that our normalization constant A is proportional to the explosion energy (and hence to maximum attainable energy) unlike the explosion energy independent normalization constant as adopted in [\cite{kac05}]. 

\subsection{Mass composition}
According to the proposed model, cosmic rays below and just above the $E_{max}^{min}$ are produced in BH formation explosions of comparable progenitor's mass. Hence there should not be any abrupt change in mass composition through the $E_{max}^{min}$. In this model higher energy particles originate from the sources of heavier progenitor. Since BH is the last stage of evolution of massive stellar objects, the composition is unlikely to change much for BHs of heavier progenitors. Therefore, the resulting composition of accelerated cosmic rays in the proposed model is expected to remain almost unaltered with energy or may become slightly heavier at higher energies. 

\section{Discussion}
We shall compare now the outcomes of the proposed model against the observational features of cosmic rays. \\
The conventional estimate of cosmic ray luminosity of our galaxy is $\sim 5 \times 10^{40} \; erg \; s^{-1}$. As shown the previous section, the proposed model yields a cosmic ray luminosity equals to $3\rho_{BH} \zeta \times 10^{43}$ ergs/s. Typically $\zeta$ ranges from 0.01 to 0.1 whereas $\rho_{BH}$ is around 0.5 \cite{cla15}. Therefore, the power from the BH producing explosions in the galaxy satisfies the power requirement for accelerating all galactic cosmic rays. Note that with the rate of occurrence one per thirty years and the average energy released in each supernova explosion around $10^{51}$ ergs, SNRs satisfy the energy budget of observed cosmic rays (hence favored as main source of cosmic rays) provided energy conversion efficiency parameter $\zeta$ is relatively higher, around 0.1 to 0.2.  \\
The maximum energy that can be attained by a cosmic ray particle in relativistic shock acceleration under the framework of the proposed model varies from source to source (of the same kind). Because of the relativistic effect (through the Lorentz factor) and owing to the much larger explosion energy, the minimum $E_{max}$ for cosmic rays is found equal to few PeV as shown in the previous section which can be identified as the knee energy. Interestingly the minimum $E_{max}$ for AGN is about 3 PeV \cite{ste91}. Whereas the maximum $E_{max}$ is found to exceed even the ankle energy. So the maximum attainable energy requirement is satisfied in a generic way. In contrast the maximum energy that can be attained by a cosmic ray particle in an ordinary SNR is 0.3 PeV which is falling short of the knee by about one order of magnitude unless the idea of magnetic amplification is invoked. Even with magnetic amplification it is difficult to exceed 100 PeV and thereby a new source of unknown nature is required between 100 PeV and the ankle energy.   \\
Since the proposed model relies on the standard shock acceleration theory, the overall cosmic ray production spectrum will follow power law behavior with spectral index equals to -2.2. Due to diffusive propagation of cosmic rays through the interstellar medium the slope of the  spectrum observed at the Earth should be steepen to $\sim 2.7$ till the knee of the spectrum and the knee should be a sharp one as observed. Above the knee the spectrum will be modified by $0.35$ due to the distribution of $E_{max}$ as demonstrated in section 3.3. Thus the proposed model explains well the observed features of energy spectrum of primary cosmic rays. \\
In respect to mass composition of cosmic rays, particularly above the knee energy,the model predicted composition is similar to that of the cannonball model but different than the prediction of supernova model of cosmic ray origin. \\
Very recent findings by the KASCADE-GRANDE collaboration about the existence of a Fe-knee around 80 PeV along with the heavier dominated composition scenario \cite{ape13, ape12, ape11} together with earlier results of KASCADE experiment for a proton knee at $3$ PeV \cite{ape10} do not support the composition picture predicted by the proposed model.  Importantly in the overlapping energy region around $1$ EeV, the composition scenario inferred from the KASCADE-GRANDE or ICETOP findings of mixed composition with nearly same contribution from proton and Iron \cite{ape10} is not in agreement with a proton dominated chemical composition as emerged from the observations of Pierre Auger Observatory \cite{abr10} , HiRes \cite{abb08, abb04} and Telescope Array \cite{tsu11, jui12}. This only shows the difficulty in estimating primary masses from air shower experiments that rests on comparisons of data to EAS simulations with the latter requires hadronic interaction models as input which are still uncertain to a large extent at present. Moreover, the uniqueness of solutions of  primary energy spectra in the knee region from EAS data is also questioned \cite{ter07} . It is expected that the mass composition scenario predicted by the present model will motivate newer experiments, exploiting both e/m and optical techniques, to establish unambiguous  cosmic ray mass composition in the knee region and in particular to confirm the KASCADE-Grande results including the Fe-knee.  \\ 
An important question is to identify the sources or more precisely identifying the gravitational explosions those lead to formation of black holes. The viable galactic sources resulting in BH formation include SN 1b/1c, hypernovae whereas GRB and AGN seem possible extragalactic sources. The observed rate of Type 1b and 1c SNe is around $10^{-3}\; yr^{-1}$ which is close to the rate of stars having mass greater than $20 M_{\odot}$. Radio observations suggest that about $5 \%$ SN 1b/1c can be produced in GRBs \cite{ber03}. Earlier Sveshnikova demonstrated that hypernovae can satisfy the power requirement for accelerating all galactic cosmic rays \cite{sve04} assuming the rate of hypernovae is about $10^{-4}$ $y^{-1}$. Extragalactic origin of cosmic rays is usually considered as unlikely on the energetic grounds. However, such a problem can be circumvented by employing flux trapping hypothesis as proposed in \cite{pla98, bur62}. Hence the possibility of GRB/AGN as the sole kind of dominate source of cosmic ray source cannot be totally ruled out from energetic consideration.    
\section{Conclusion}
In summary, the proposed speculative BH based model of origin of cosmic rays can account all the major observed features about cosmic rays without any serious contradiction to observational results. The knee of the energy spectrum has been ascribed as the consequence of the mass distribution of progenitor of cosmic ray source. Such a philosophy seems applicable to the Cannonball model of cosmic ray origin replacing the original proposal of second order Fermi acceleration of cosmic rays by Cannonballs of other SN explosions as the cause of spectral steepening above the knee \cite{dar99, pla02, dar05, der05}. Precise measurement of primary mass composition can be used to discriminate the proposed model from most of the standard prevailing models of cosmic ray knee. No definite  cosmic ray sources could be identified at this stage within the framework of the proposed model which would be an important future task for further development of the proposed model.

{\bf Acknowledgement}
The authors are thankful to an anonymous reviewer for insightful comments and suggestions that help us to improve the manuscript. AB thanks Professors C. L. Fryer and S. E. Woosley for helpful discussions. This work is partly supported by the Department of Science and Technology (Govt. of India) under the grant no. SR/S2/HEP-14/2007.

\label{lastpage}


\begin{thebibliography}{99}

\bibitem[Aartsen, et al(2013)]{aar13}  Aartsen, M.G. et al. (ICETOP Coll.), 2013 Phys. Rev. D 88, 042004 
\bibitem[Abbasi et al(2008)]{abb08} Abbasi, R. et al., 2008, Phys. Rev. Lett. 100, 101101 
\bibitem[Abbasi et al(2004)]{abb04} Abbasi, R. et al., 2004, Phys. Rev. Lett. 92, 151101
\bibitem[Abraham(2010)]{abr10} Abraham, J., 2010, Phys. Rev. Lett. 104, 091101 
\bibitem[Aglietta et al(1990)]{agl90} Aglietta, M. et al., 1990 Nucl. Phys. (Proc. Supl.) 14B, 193 
\bibitem[Ahlen et al(1992)]{ahl92} Ahlen, S. et al., 1992 Phys. Rev. D 46, 4836 
\bibitem[Aloisio et al(2012)]{alo12} Aloisio, R., Berezinsky, V. and Gazizov, A., 2012 Astropart. Phys. 39-40, 129
\bibitem[Amato(2014)]{ama14} Amato, E., 2014 Int. J. Mod. Phys. D, 23, 1430013
\bibitem[Apel et al(2010)]{ape10}  Apel, W.-D. et al.(KASCADE Collaboration), 2010 Astropart. Phys. 31, 86
\bibitem[Apel et al(2011)]{ape11}  Apel, W.-D. et al.(KASCADE Collaboration), 2011 Phys. Rev. Lett. 107, 171104 
\bibitem[Apel et al(2012)]{ape12}  Apel, W.-D. et al.(KASCADE Collaboration), 2012 Astropart. Phys. 36, 183
\bibitem[Apel et al(2013)]{ape13}  Apel, W.-D. et al.(KASCADE Collaboration), 2013 Phys. Rev. D87, 081101
\bibitem[Arqueros et al(2000)]{arq00} Arqueros, F. et al., 2000, Astron. Astrophys. 359, 682
\bibitem[Bakatanov et al(1999)]{bak99} Bakatanov, V. N. et al., 1999 Astropart. Phys. 12, 19
\bibitem[Berezhko(1999)]{ber99} Berezhko, E.G. and Ksenofontov, L.T., 1999, JETP 89, 391 
\bibitem[Berger et al(2003)]{ber03} Berger, E. et al., 2003 \apj 599, 408 
\bibitem[Bhadra \& Sanyal(2005)]{bha05a} Bhadra, A. and Sanyal, S., 2005, in Proc. 29th Int. Cosmic Ray Conf., 6, 137
\bibitem[Bhadra, A(2005)]{bha05} Bhadra, A., 2005 Proc. 29th Int. Cosmic Ray Conf. 3, 117
\bibitem[Biermann(1993)]{bie93} Biermann, P. L., 1993, Astron. Astrophys. 271, 649
\bibitem[Blake and Nash(1995)]{bla95} Blake, P. R. and Nash, M.F., 1995 J. Phys. G: Nucl. Part. Phys. 21, 1731
\bibitem[Blake and Nash(1998)]{bla98} Blake, P. R. and Nash, M.F., 1998 J. Phys. G: Nucl. Part. Phys. 24, 217 
\bibitem[Blasi(2014)]{bla14} Blasi, P. 2014 R. Physique 15, 329
\bibitem[Boothby et al(1997)]{boo97} Boothby, K., et al., 1997 Astrophys.J 491, L35 
\bibitem[Burbidge(1962)]{bur62} Burbidge, G., 1962 Prog. Theor. Phys. 27, 999 
\bibitem[Candia et al (2002)]{can02} Candia, J., Roulet, E., and Epele, N. 2002, JHEP 0212, 033
\bibitem[Candia et al(2002)]{can02a} Candia, J., Epele, L. N. and Roulet, E. 2002 Astropart. Phys. 17, 23 
\bibitem[Chakrabarty et al(1998)]{cha98} Chakrabarty, C. et al., 1998 IL Nuovo Cim. 21, 215 
\bibitem[Christodoulou and Ruffini(1971)]{chr71} Christodoulou, D. and Ruffini, R. 1971 Phys. Rev. D 4, 3552
\bibitem[Cherev et al(2004)]{che04} Cherev, D. et al., 2004, astro-ph/0411139
\bibitem[Clausen et al(2015)]{cla15} Clausen, D., Piro, A.L. and Ott, C. D., 2015 Astrophys. J. 799, 190 
\bibitem[Danilova et al(1995)]{dan95} Danilova, E. V. et al., 1995 Proc. 24th Int. Cosmic Ray Conf, 1, 286 
\bibitem[Dar and Plaga(1999)]{dar99} Dar, A. and Plaga, R. 1999 Astron. Astrophys. 349, 259
\bibitem[Dar(2005)]{dar05} Dar, A., 2005 Nuovo Cim. B 120, 767
\bibitem[De Rujula(2005)]{der05} De Rujula, A., 2005 Int. J. Mod. Phys.A20, 6562
\bibitem[Dickinson et al(1999)]{dic99} Dickinson, J. E. et al., 1999, in Proc. 26th Int Cosmic Ray Conf. 3, 136
\bibitem[Efimov et al(1991)]{efi91} Efimov, N. N. et al., 1991, Proc. Int. Symp. astrophysical aspects of most energetic cosmic rays, Ed. M. Nagano, and F. Takahara, p 20 
\bibitem[Erlykin and Wolfendale(1997)]{erl97} Erlykin, A.D. and Wolfendale, A.W., 1997 J. Phys. G: Nucl. Part. Phys. 23, 979
\bibitem[Erlykin et al(2011)]{erl11} Erlykin, A. D., Martirosov, R. and Wolfendale, A. 2011, CERN Courier, 51, 21
\bibitem[Fichtel and Linsley(1986)]{fic86} Fichtel, C.E. and Linsley, J. 1986, \apj 300, 474
\bibitem[Fomin et al(1996)]{fom96} Fomin, Y. et al., 1996 J. Phys. G: Nucl. Part. Phys. 22, 1839
\bibitem[Fowler et al(2001)]{fow01} Fowler,J. F. et al., 2001, Astropart. Phys. 15, 49 
\bibitem[Fryer(1999)]{fry99} Fryer, C. L. 1999 \apj 522, 413
\bibitem[Fryer and Heger(2000)]{fry00} Fryer, C. L. and Heger, A.  2000 \apj 541, 1033
\bibitem[Fryer(2003)]{fry03} Fryer, C. L. 2003 Class. Quantum Grav. 20, S73
\bibitem[Giacinti et al (2014)]{gia14} Giacinti, G., Kachelrie, M., and Semikoz, D. V. 2014, Phys. Rev. D 90, 041302(R)
\bibitem[Ginzburg and Syrovatskii(1964)]{gin64} Ginzburg, V. L. and Syrovatskii, S. I. 1964, The Origin of Cosmic Rays, Macmillan, NewYork
\bibitem[Glasmacher et al(1999)]{gla99}  Glasmacher, M. et al.(CASA-MIA Collab.), 1999 Astropart. Phys. 12, 19
\bibitem[Haungs(2011)]{hau11}  Haungs, A., 2011 Space Sc. Trans, 7, 295 
\bibitem[Heger and Fryer(2003)]{heg03} Heger, A. and Fryer, C. L. 2003 \apj 591, 288
\bibitem[Hillas(1979)]{hil79} Hillas, A. M., 1979 Proc. 16th Int. Cosmic Ray Conf., 8, 7
\bibitem[Hillas(1984)]{hil84} Hillas, A. M. 1984, Ann. Rev. Astron. Astrophys. 22, 425  

\bibitem[Horandel(2013)]{hor13} Horandel, J.R., 1993, AIP Conf. Proc., 1516, 185
\bibitem[Hu et al(2009)]{hu09} Hu, H-B., et al, 2009, \apj, 700, L170
\bibitem[Jokipii(1986)]{jok86} Jokipii, J.R.  and Morfill, G.E. 1986, \apj 312, 170 
\bibitem[Jui(2012)]{jui12} Jui, C. C., 2012, J. Phys. Conf. Ser. 404, 012037 
\bibitem[Kachelriess and Semikoz(2005)]{kac05} Kachelriess, M. and Semikoz, D. V. 2005 arXiv:astro-ph/0510188 
\bibitem[Karakula and Tkaczyk(1993)]{kar93} Karakula, S. and Tkaczyk, W., 1993 Astropart. Phys. 1, 229 
\bibitem[Karle et al(1995)]{kar95} Karle, A. et al., 1995, Astropart. Phys. 3, 321

\bibitem[Kasahara et al(1997)]{kas97} Kasahara, S. et al., 1997 Phys. Rev. D 55, 5282
\bibitem[Kroupa(2002)]{kro02} Kroupa, P. 2002 Science 295, 82
\bibitem[Kobayakawa (2002)]{kob02} Kobayakawa, K. et al., 2002, Phys. Rev. D 66, 083004 
\bibitem[Kulikov and Khristiansen (1959)]{kul59} Kulikov, G. V. and Khristiansen, G. B. 1959 JETP, 35, 441.
\bibitem[Lithwick and Sari(2001)]{lit01} Lithwick, Y. and Sari, R. 2001 \apj 555, 540 
\bibitem[Longley et al.(1995)]{lon95} Longley, N. et al., 1995 Phys. Rev. D 52, 2760
\bibitem[Massey et al(1995)]{mas95} Massey, P. et al. 1995 \apj 454, 151
\bibitem[Navarra et al(1998)]{nav98} Navarra, G. et al. (EAS TOP Collab.), 1998 Nucl. Phys. (Proc. Supl.) 12, 1
\bibitem[Nikolsky and Romachin(2000)]{nik00} Nikolsky, S. I. and Romachin, V. A. 2000 Physics of Atomic Nuclei, 63, 1799

\bibitem[Plaga(1998)]{pla98} Plaga, R. 2001 Astron. Astrophys.330, 833 
\bibitem[Plaga(2002)]{pla02} Plaga, R., 2002 New Astronomy, 7, 317
\bibitem[Ptuskin(1993)]{ptu93} Ptuskin, V. S., 1993, Astron. Astrophys. 268, 726
\bibitem[Racusin et al(2011)]{rac11} Racusin, J. L. et al., 2011 \apj 738, 138 
\bibitem[Ruffini and Vitagliano(2003)]{ruf03} Ruffini, R. and Vitagliano, L., 2003 Int. J. Mod. Phys. D, 12, 121
\bibitem[Salpeter(1955)]{sal55} Salpeter, E. E., 1955 \apj 121, 161
\bibitem[Stanev et al(1993)]{sta93} Stanev T., et al., 1993, Astron. Astrophys. 274, 902 
\bibitem[Stecker et al(1991)]{ste91} Stecker, F. W., Done, C, Salamon, M. H. and Sommers, P., 1991 Phys. Rev. Letts 66, 2697 
\bibitem[Sveshnikova(2003)]{sve03} Sveshnikova, L. G., 2003 Astron. Astroph. 409, 799
\bibitem[Sveshnikova(2004)]{sve04} Sveshnikova, L. G., 2004 Astron. Letts. 30, 41
\bibitem[Swordy et al.(2001)]{swo01} Swordy S. P. et al., 2001, Astropart. Phys. 13, 137 

\bibitem[Ter-Antonyan(2007)]{ter07} Ter-Antonyan, S., 2007, Astropart. Phys. 28, 321
\bibitem[Ter-Antonyan(2014)]{ter14} Ter-Antonyan, S., 2014 Phys. Rev. D. 89, 123003
\bibitem[Tsunesada(2011)]{tsu11} Tsunesada, Y., Preprint arXiv:1111.2507

\bibitem[Wdowczyk and Wolfendale(1984)]{wdo84} Wdowczyk, J. and Wolfendale, A.W. 1984, J. Phys. G, 10, 1453 
\bibitem[Woosley(1993)]{woo93} Woosley, S. E. 1993 \apj 405, 273
\bibitem[Zou, Fan and Piran(2011)]{zou11} Zou, Y.-C., Fan, Y.-Z., Piran T., 2011, \apj 726, L2  
\end{thebibliography}
\end{document}